\documentclass[12pt]{article}

\usepackage{paralist}
\usepackage{wrapfig}
\usepackage{graphicx}
\usepackage[none]{hyphenat}
\usepackage[english]{babel}
\usepackage[utf8]{inputenc}
\usepackage[T1]{fontenc}
\usepackage{tikz-cd}
\usepackage{epsfig}

\usepackage{epstopdf}

\usepackage{parskip}

\usepackage{authblk}

\usepackage{caption}

\usepackage{csquotes}
\usepackage{mathtools,amssymb,amsfonts,amsthm}

\usepackage[breaklinks,colorlinks=true]{hyperref}
\hypersetup{urlcolor=blue, citecolor=red}
\usepackage[capitalise]{cleveref} 

\theoremstyle{plain}
\newtheorem{theorem}{Theorem}
\newtheorem*{theorem*}{Theorem}
\newtheorem{lemma}[theorem]{Lemma}
\newtheorem*{lemma*}{Lemma}
\newtheorem{proposition}{Proposition}
\newtheorem*{proposition*}{Proposition}
\newtheorem{corollary}{Corollary}
\newtheorem*{corollary*}{Corollary}

\theoremstyle{definition}
\newtheorem{definition}{Definition}
\newtheorem*{definition*}{Definition}

\newtheorem*{example*}{Example}

\crefname{theorem}{Theorem}{Theorems}
\Crefname{theorem}{Theorem}{Theorems}
\crefname{lemma}{Lemma}{Lemmas}
\Crefname{lemma}{Lemma}{Lemmas}
\crefname{proposition}{Proposition}{Propositions}
\Crefname{Prop}{Proposition}{Propositions}
\crefname{corollary}{Corollary}{Corollaries}
\Crefname{corollary}{Corollary}{Corollaries}
\crefname{definition}{Definition}{Definitions}
\Crefname{definition}{Definition}{Definitions}
\crefname{example}{Example}{Examples}
\Crefname{example}{Example}{Examples}
\crefname{theorem}{Theorem}{Theorems}

\newtheorem*{exercise*}{Exercise}
\crefname{exercise}{exercise}{exercises}
\Crefname{exercise}{Exercise}{Exercises}

\newcommand{\pdv}[2]{\frac{\partial #1}{\partial #2}}

\DeclarePairedDelimiter{\set}{\lbrace}{\rbrace}

\newcommand{\RR}{\mathbb{R}}
\newcommand{\dd}{\mathrm{d}}

\newcommand{\Reeb}{\mathcal{R}}
\newcommand*{\ann}[1]{{#1}^{\circ}} 

\newcommand*{\contr}[1]{\iota_{#1}}

\newcommand*{\orth}[1]{{#1}^{\bot}}
\newcommand*{\orthL}[1]{\prescript{\bot}{}{#1}}

\hypersetup{urlcolor=blue, citecolor=red}

\title{The contact Eden bracket and the evolution of observables}

\author{
\begin{center}

V{\'\i}ctor M. Jim\'enez\footnote{E-mail: \href{mailto:victor.jimenez@mat.uned.es}{victor.jimenez@mat.uned.es}}
\\  Universidad Nacional de Educación a Distancia (UNED), \\
Departamento de Matemáticas Fundamentales. \\
Calle de Juan del Rosal 10, 28040, Madrid, Spain

\bigskip

Manuel de Le\'on\footnote{E-mail: \href{mailto:mdeleon@icmat.es}{mdeleon@icmat.es}}
\\ Instituto de Ciencias Matem\'aticas, Campus Cantoblanco \\
 Consejo Superior de Investigaciones Cient\'ificas
 \\
C/ Nicol\'as Cabrera, 13--15, 28049, Madrid, Spain
\\
and
\\
Real Academia de Ciencias de España.
\\
C/ Valverde, 22, 28004 Madrid, Spain.
\end{center}
}

\begin{document}
\maketitle

\begin{abstract}
In this paper we discuss nonholonomic contact Lagrangian and Hamiltonian systems, that is, systems with a kind of dissipation that are also subject to nonholonomic constraints. We introduce the so-called contact Eden bracket that allows us to obtain the evolution of any observable. Finally, we present a particular vector subspace of observables where the dynamics remain unconstrained.
\end{abstract}
\tableofcontents

\section{Introduction}
\sloppy
The study of mechanical systems with non-holonomic constraints is a classical topic of great interest in the engineering sciences. Recently, the authors have begun to study these systems in a different context, contact geometry, which allows systems with dissipation to be considered \cite{primero,segundo}. 
Indeed, contact Hamiltonian systems are experiencing a revival in recent years (see \cite{Bravetti2017,deLeon2020b}), and especially for their applications to the study of thermodynamic systems.

The equations of motion for the nonholonomic contact systems, as well as the nonholonomic brackets that adequately describe the dynamics in the constraint submanifold, were obtained from the cited papers of the authors.

This nonholonomic bracket is obtained by a convenient decomposition of the phase space, and the corresponding projections along the submanifold constraint. Moreover, in the case of mechanical Lagrangians, i.e., those coming from a Riemannian metric in configuration space (and a dissipative potential), the decomposition can be refined so that the nonholonomic bracket is expressed in a very simple and computationally tractable way using the projection $\gamma$ (denoted in this manner to maintain the notation of the original papers by Eden \cite{eden1,eden2}). The results previously known in the symplectic context \cite{CaLeDi99,deleonetal} are thus extended to the contact context.

The main results of the current paper are the following. The evolution of an observable $f$ in the non-holonomic system corresponds to the evolution of the observable $f \circ \gamma$ in the unconstrained system. And, in addition, for observables that fulfill the so-called mechanical condition, the Eden bracket just restricts to the canonical Jacobi bracket. 

The paper is structured as follows.  In Section 2 we introduce the notion of constrained contact Hamiltonian systems whereas the Lagrangian counterpart is studied in Section 3. In Section 5 we define the Eden bracket which is used in Section 5 to discuss the evolution of observables.

\section{Constrained contact Hamiltonian systems}\label{sec:contact_jacobi}

Let us consider a \textit{Hamiltonian function} as a differentiable function $H : T^{*}Q\times \mathbb{R} \rightarrow \mathbb{R}$. Consider the canonical contact form of $T^{*}Q\times \mathbb{R}$ given by
\begin{equation}
    \eta_{Q} = dz - \theta_{Q} = dz - p_{i} \dd q^{i},
\end{equation}
where $\left( q^{i}, p_{i} , z \right)$ are canonical coordinates; the corresponding Reeb vector field is $\Reeb_{Q} = \pdv{}{z}$.

There exists a vector bundle isomorphism,
\begin{equation}\label{eq:flat_iso}
    \begin{aligned}
        {\flat}_{Q} : T\left( T^{*}Q \times \mathbb{R}\right)&\to T^* \left( T^{*}Q \times \mathbb{R}\right) ,\\
         v &\mapsto \contr{v}  \dd \eta_{Q} + \eta_{Q} (v)  \eta_{Q},
    \end{aligned}
\end{equation}
whose inverse will be denoted by $\sharp_{Q}$. 
The \textit{contact Hamiltonian vector field} $X_H$ for a \textit{Hamiltonian function} $H:  T^{*}Q \times \mathbb{R} \rightarrow \mathbb{R}$, is defined by the formula,
\begin{equation}\label{eq:hamiltonian_vf_contact}
    {\flat}_{Q} (X_H) = \dd H - (\Reeb_{Q} (H) + H) \, \eta_{Q}.
\end{equation}
Then, it is locally expressed as follows,
\begin{equation}\label{Hamiltneq2}
    X_{H} =\pdv{H}{p_i}\pdv{}{q^i}  -\left(\pdv{H}{q^i}  +p_{i} \pdv{H}{z}\right) \pdv{}{p_{i}}    + \left( p_{i} \pdv{H}{p_{i}}-H\right)\pdv{}{z}
\end{equation}
Therefore, its integral curves satisfy the \textit{contact Hamiltonian equations}
\begin{equation}\label{Hamiltoneqs34}
\begin{dcases}
 \frac{\dd q^{i}}{\dd t} &=    \pdv{H}{p^{i}}\\
 \frac{\dd p_{i}}{\dd t} &=  - \pdv{H}{q^{i}}  - p_{i} \pdv{H}{z}\\
 \frac{\dd z}{\dd t} &=   p_{i} \pdv{H}{p_{i}}-H
 \end{dcases}
\end{equation}

The contact structure $\eta_{Q}$ induce the so-called \textit{contact bracket}, which is a Jacobi bracket of functions, as follows,
\begin{equation}\label{ContNH15}
  \set{f,g} = -\dd \eta_{Q}(\sharp_{Q}\left(\dd f\right), \sharp_{Q}\left(\dd g\right)) - f \Reeb_{Q}(g) + g \Reeb_{Q}(f), \ \forall f,g \in T^{*}Q \times \mathbb{R}.
\end{equation}

\begin{theorem}\label{genesteteorema23}
  The contact bracket (\ref{ContNH15}) provides the evolution of the observables, that is,
    \begin{equation}
      X_{H}(f) = \set{H,f} - f \Reeb_{Q} (H).
    \end{equation}
\end{theorem}

Let us define the following $2-$tensor,
\begin{equation}
    \omega_Q = \dd \eta_Q + \eta_Q \otimes \eta_Q,
\end{equation}
so that ${\flat}_Q$ is the contraction of $\omega_Q$ (see \cref{eq:flat_iso}). 
\begin{definition}
Let be $\mathcal{S}$ a distribution on $ T^{*}Q\times \mathbb{R}$. Then, we may define the following notions of complement with respect to $\omega$:
\begin{equation}
    \begin{aligned}
        \orth{\mathcal{S}}  &=
          \set{v \in T\left( T^{*}Q\times \mathbb{R}\right) \mid \omega_{Q}(w,v) = {\flat}_{Q}(w)(v) = 0,\, \forall w\in \mathcal{S}}
          = \ann{({\flat}_{Q}(\mathcal{S}))},\\
        \orthL{\mathcal{S}} &=
          \set{v \in T\left( T^{*}Q\times \mathbb{R}\right)\mid \omega_{Q}(v,w) = 0, \forall w \in \mathcal{S}} = \flat_{Q}^{-1}(\ann{\mathcal{S}}).
    \end{aligned}
\end{equation} 
$\orth{\mathcal{S}}$ and $\orthL{\mathcal{S}}$ will be called \textit{right} and \textit{left orthogonal complements respectively}.
\end{definition}

\section{Constrained Lagrangian systems}

Let $L$ be a Lagrangian  of \textit{mechanical type}, i.e., there exists a Riemannian metric $g$ on $Q$, such that,
$$L \left(v_{q},z\right) \ = \ \frac{1}{2}g\left( v_{q},v_{q}\right) - V\left( q,z\right), \ \forall \left(v_{q},z\right) \in T_{q}Q\times \mathbb{R}.$$
Here, $T \left(v_{q}\right)= \frac{1}{2}g\left( v_{q},v_{q}\right)$ represent the \textit{kinetic energy}, and $V$ is the \textit{potential energy}. Then, the image by the Legendre transformation $FL$ is given by,
$$FL \left( \mathcal{D}\times \mathbb{R}\right) = \flat_{g}\left( \mathcal{D}\right) \times \mathbb{R},$$
where $\flat_{g}$ is the \textit{musical isomorphism} associated to $g$. Denote by $\mathcal{M} = \flat_{g}\left( \mathcal{D}\right) = \ann{\mathcal{D}^{{\perp}_{g}}}$ the vector subbundle of $T^{*}Q$ isomorphic to $\mathcal{D}$, where $\cdot^{\perp_{g}}$ defines the orthogonal complement respect to the Riemannian metric $g$. We will denote the inverse of $\flat_{g}$ by $\sharp_{g}$, i.e., $\flat_{g}^{-1} = \sharp_{g}$. Then, the associated Hamiltonian function is given by,
\begin{equation}\label{ContNH14}
    H \left( q^{i} , p_{i} , z  \right) = \frac{1}{2} g^{i,j}p_{i}p_{j} + V\left( q^{i} , z\right) ,
\end{equation}
where $g^{i,j}$ are the entries of the inverse matrix of $\left( g_{i,j} = g\left( \frac{\partial}{\partial q^{i}} , \frac{\partial}{\partial q^{j}}\right)\right)$.

\begin{theorem}[\cite{segundo}]\label{18.3}
Let be a Hamiltonian function $H: T^{*}Q\times \mathbb{R} \rightarrow \mathbb{R}$, and a constraint manifold $\mathcal{M} \subseteq T^{*}Q$ as above. Let $X$ be a vector field on $T^{*}Q \times \RR$ satisfying the equations
\begin{equation}\label{6.3}
 \begin{dcases}
        \flat_Q\left(X\right) - \dd H + \left(H + \Reeb_Q\left(H\right)\right)\eta_Q \in \ann{\mathfrak{D}^{l}} \\
        X_{\vert \mathcal{M} \times \mathbb{R}} \in  \mathfrak{X}\left(\mathcal{M} \times \mathbb{R} \right).
    \end{dcases}
\end{equation}
where
$$ \ann{\mathfrak{D}^{l}} = \left( \pi_{Q,\mathbb{R}}\right)^{*} \flat_{g}\left( \ann{\mathcal{M}}\right)$$
Then, the integral curves of $X$ are solutions of the following constrained Hamiltonian Herglotz equations,
\begin{align}
    \begin{cases}\label{eq:Hamiltonnonholonomic_herglotz_eqs_coordsMechatype}
         \frac{\dd q^{i}}{\dd t} &=   \pdv{H}{p^{i}}\\
 \frac{\dd p_{i}}{\dd t} &=  - \pdv{H}{q^{i}}  -p_{i} \pdv{H}{z} - \lambda_a\Phi^a_i\\
 \frac{\dd z}{\dd t} &=   p_{i} \pdv{H}{p_{i}}-H\\
        \varphi^{a}\left(\flat_{g}\left(\dot{\xi}\right)\right) & = 0, \ \forall a
    \end{cases}
\end{align}
where $\ann{\mathcal{D}}$ is (locally) generated by the basis of $1-$forms $\Phi^{a}_{i}\dd q^{i}$.
\end{theorem}
In this case, the solution of \cref{6.3} is called \textit{constrained Hamiltonian vector field}, and will be denoted by $X_{H, \mathcal{M}}$.

\section{The contact Eden bracket}

Assume that we have a constrained contact Hamiltonian system on $T^{*}Q \times \mathbb{R}$ given by a Hamiltonian $H$ and a vector subbundle $\mathcal{M}$ of the cotangent bundle $T^{*}Q$; $H$ corresponds to a Lagrangian function of mechanical type on $TQ \times \mathbb{R}$.

Notice that,
$$TQ = \mathcal{D}\oplus \mathcal{D}^{\perp_{g}},$$
where $\mathcal{D}= \sharp_{g}\left( \mathcal{M}\right)$. Then, by applying $\flat_{g}$, we have that
$$T^{*}Q = \mathcal{M}\oplus \ann{\mathcal{D}}$$
In particular,
\begin{equation}\label{eq2131234}
T^{*}Q\times \mathbb{R} = \left(\mathcal{M}\times \mathbb{R}\right)\oplus \ann{\mathcal{D}}
\end{equation}
Let us consider the projection $\gamma: T^{*}Q\times \mathbb{R} \rightarrow \mathcal{M}\times \mathbb{R}$. Then, it turns natural to define the following bracket of functions,
\begin{equation}\label{56}
\set{f,g}_{E} = \set{f \circ \gamma,g\circ \gamma}_{\vert \mathcal{M}\times \mathbb{R}}
\end{equation}
for all $f,g \in \mathcal{C}^{\infty}\left(\mathcal{M}\times \mathbb{R}\right)$. This bracket is called \textbf{contact Eden bracket} following the spirit of \cite{eden1,eden2}, and coincides with the other known nonholonomic brackets defined in \cite{primero,segundo}. Let us assume that the projection $\gamma$ is locally given by the formula,
$$ \gamma \left( q^{i}, p_{i} , t\right) = \left(q^{i}, \gamma^{l}_{i}p_{l} , t \right),$$
i.e., $\left(\gamma^{i}_{j}\right)$ is the projection matrix associated to $\gamma$. Therefore,
\begin{equation}
\left( q^{i}, p_{i} , t\right) \in  \mathcal{M} \times\RR \ \Leftrightarrow \  p_{i} =  \gamma^{l}_{i}p_{l}, \ \forall i
\end{equation}
On the other hand, $v \in \orthL{\mathfrak{D}^{l}} = \sharp_{Q}\left(\ann{\mathfrak{D}^{l}}\right) $ if, and only if, there are local functions $\lambda_{a}$ such that,
\begin{equation}\label{ContNH1}
    \contr{v}\dd \eta_{Q} + \eta_{Q}\left( v \right) \eta_{Q} = \lambda_{a} \Phi^{a}_{i} \dd q^{i}.
\end{equation}
Notice that, $\pdv{}{p_i}, \pdv{}{z} \in   \mathfrak{D}^{l}$ and, thus, $v\left( q^{i}\right)=0=v\left( z\right)$.
Therefore, using \cref{ContNH1}, we have that $v \in \orthL{\mathfrak{D}^{l}} = \sharp_{Q}\left(\ann{\mathfrak{D}^{l}}\right) $ if, and only if, there are local functions $\lambda_{a}$ such that,
\begin{equation}\label{ContNH2}
   v\left( q^{i}\right)=0=v\left( z\right), \ \ \  v\left(p_{i}\right) = \lambda_{a} \Phi^{a}_{i},
\end{equation}

Since $L$ is of mechanical type, we have
\begin{equation}\label{ContNH3}
 \Phi^{a}_{i} \pdv{H}{p_i} = 0,
\end{equation}
evaluated at elements of $\mathcal{M}$. In other words, the vector $\pdv{H}{p_i}\pdv{}{q_i}$ is tangent to $\mathcal{D}$. Then,
\begin{align*}
    \flat_{Q}\left(\pdv{H}{p_i}\pdv{}{q_i}\right) &=\pdv{H}{p_i}\flat_{Q}\left(\pdv{}{q_i}\right)\\
    &= \pdv{H}{p_i}g_{i,j} \dd q^{j} \in \mathcal{M}\\
\end{align*}
As a consecuence, we have that,
$$\gamma^{l}_{i}\ g_{k,l}\pdv{H}{p_k} = g_{k,i}\pdv{H}{p_k}, \ \forall i.$$
Due to $\gamma$ is an orthogonal projection, the matrices $\left( \gamma^{j}_{i}\right)$ and $\left(g_{i,j}\right)$ are symmetric and, therefore,
$$g_{l,i}\gamma^{k}_{l}\pdv{H}{p_k} = g_{k,i}\pdv{H}{p_k}, \ \forall i.$$
Taking into account that $\left(g_{i,j}\right)$ is a regular matrix, we have proved that, for elements at $\mathcal{M}$,
\begin{equation}\label{ContNH4}
\gamma^{k}_{i}\pdv{H}{p_k} = \pdv{H}{p_i}, \ \forall i
\end{equation}
In other terms,
\begin{equation}\label{ContNH5}
\pdv{\left(H \circ \gamma\right)}{p_i} = \pdv{H}{p_i}, \ \forall i.
\end{equation}
On the other hand, it is obvious that, over $\mathcal{M}\times \mathbb{R}$,
\begin{equation}\label{ContNH6}
\pdv{\left(H \circ \gamma\right)}{z} = \pdv{H}{z}, \ \forall i.
\end{equation}

Let us now fix a local section  of $\mathcal{M}$ given by $\alpha = \alpha^{i}\dd q^{i}$. Then, \cref{ContNH4} implies that,
$$\pdv{\gamma^{k}_{j}}{q^{i}}\pdv{H}{p_k} + \gamma^{l}_{j}\left[\pdv{\alpha^{m}}{q^{i}} \frac{\partial^{2}H}{\partial p_{m}\partial p_{l}} + \frac{\partial^{2}H}{\partial q^{i}\partial p_{l}}\right] = \pdv{\alpha^{m}}{q^{i}} \frac{\partial^{2}H}{\partial p_{m}\partial p_{j}} + \frac{\partial^{2}H}{\partial q^{i}\partial p_{j}}, \ \forall i,j
$$
Then,
\begin{align*}
p_{j}\pdv{\gamma^{k}_{j}}{q^{i}}\pdv{H}{p_k} &=  p_{j}\left[\pdv{\alpha^{m}}{q^{i}} \frac{\partial^{2}H}{\partial p_{m}\partial p_{j}} + \frac{\partial^{2}H}{\partial q^{i}\partial p_{j}}\right] - p_{j}\gamma^{l}_{j}\left[\pdv{\alpha^{m}}{q^{i}} \frac{\partial^{2}H}{\partial p_{m}\partial p_{l}} + \frac{\partial^{2}H}{\partial q^{i}\partial p_{l}}\right]\\
&=0
\end{align*}
The last equality follows from the fact that we are evaluating at a elements of $\mathcal{M}$ and $\left(\gamma_{i}^{j}\right)$ is a symmetric matrix. Thus,
\begin{equation}\label{ContNH7}
 p_{j}\pdv{\gamma^{k}_{j}}{q^{i}}\pdv{H}{p_k} = 0, \ \forall i   
\end{equation}
So,
\begin{align*}
\pdv{\left(H \circ \gamma\right)}{q^{i}} &= \pdv{H}{q^{i}} +  \pdv{\gamma^{j}_{k}}{q^{i}}p_{j}\pdv{H}{p_k}\\
&=\pdv{H}{q^{i}},
\end{align*}
i.e., for all $i$,
\begin{equation}\label{ContNH8}
\pdv{\left(H \circ \gamma\right)}{q^{i}} =\pdv{H}{q^{i}}. 
\end{equation}
Therefore, using \cref{ContNH5}, \cref{ContNH6}, and \cref{ContNH8}, we have that,
\begin{equation}\label{ContNH9}
    \left(\dd H\right)_{\mathcal{M}\times \mathbb{R}} =  \left(\dd \left(H\circ \gamma\right)\right)_{\mathcal{M}\times \mathbb{R}}
\end{equation}
\begin{lemma}\label{ContNH19}
${X_{H}}_{\vert \mathcal{M} \times\RR}  = {X_{H \circ \gamma}}_{\vert \mathcal{M} \times\RR}$. 

\end{lemma}

\begin{theorem}[\cite{segundo}]\label{ContNH10}
$T\gamma \left( {X_{H}}_{\vert \mathcal{M} \times\RR} \right) = X_{H , \mathcal{M}}.$

\end{theorem}
Thus, as an immediate corollary of \cref{ContNH10} and \cref{ContNH9}, we deduce
\begin{corollary}\label{ContNH12}
Let $X_{H , \mathcal{M}}$ be the constrained Hamiltonian vector field. Then,
$$T\gamma \left( {X_{H}}_{\vert \mathcal{M} \times\RR} \right) = X_{H , \mathcal{M}} =  T\gamma\left({X_{H \circ \gamma}}_{\vert \mathcal{M}\times \mathbb{R}}\right).$$
\end{corollary}

\begin{theorem}\label{ContNH11}
  The contact Eden bracket has the following properties:
  \begin{enumerate}
    \item Any function $g$ on $T^{*}Q \times \mathbb{R}$ that is constant on $\mathcal{M}\times \mathbb{R}$ is a Casimir, i.e.,
    \[\set{f,g}_{E} = 0, \ \forall f \in \mathcal{C}^{\infty} \left( T^{*}Q \times \RR \right)\]
    \item The bracket provides the evolution of the observables, that is,
    \begin{equation}
      X_{H , \mathcal{M}}(f) = \set{H,f}_{E} - f \Reeb_{H , \mathcal{M}} (H).
    \end{equation}
  \end{enumerate}
\end{theorem}

Notice that, in particular, all the functions $\varphi^{a}$ defining $\mathcal{M}$ are Casimir's.\\
It is also remarkable that, using the statement \textit{1.} in \cref{ContNH11}, the contact Eden bracket may be restricted to functions on $M$. \textit{Thus, from now on, we will refer to the nonholonomic bracket as the restriction of $\set{\cdot , \cdot }_{E}$ to functions on $M$.}

\section{Evolution of observables}

Let us now focus on the the evolution of the observables. In particular, due to \cref{ContNH11}, we have that,
    \begin{equation}
      X_{H , \mathcal{M}}(f) = \set{H,f}_{E} - f \Reeb_{H , \mathcal{M}} (H).
    \end{equation}
Nevertheless, using \cref{ContNH12}, we may prove the following result.

\begin{proposition}
The contact Eden bracket $\set{\cdot, \cdot}_{E}$ satisfies the following properties:
\begin{enumerate}[i)]
\item Let $f$ be a function on $T^{*}Q \times \mathbb{R}$. Then,
$$ X_{H , \mathcal{M}}(f) = X_{H }(f \circ \gamma) $$
    \item For any function $f$ on $T^{*}Q \times \mathbb{R}$, we have that,
    \[\set{H,f}_{E} = \set{H,f \circ \gamma} . \]
\end{enumerate}
\end{proposition}

Thus, we have proven that the evolution of an observable $f$ in the non-holonomic system corresponds to the evolution of the observable $f \circ \gamma$ in the unconstrained system. In other words, the evolution of an observable in the non-holonomic system can be analyzed through the evolution of a corresponding observable in the system without constraints. In particular, over $\mathcal{M}\times \mathbb{R}$, we have
$$ X_{H , \mathcal{M}}(f) = X_{H }(f \circ \gamma) = \set{H,f \circ \gamma} - f \Reeb \left( H \right), \ \forall  f \in \mathcal{C}^{\infty}\left( T^{*}Q \times \mathbb{R}\right).$$

Next, we will consider a family of observables that fulfill a condition called the mechanical condition, that is,
$$
\dd f_{\vert M} \in \ann{\orthL{\mathfrak{D}^{l}}}.
$$

Thus, if we denote the vector subspace of observables satisfying the mentioned mechanical condition by $\mathcal{C}_{\mathcal{M}}^{\infty}\left( T^{*}Q \times \mathbb{R}\right)$, we have the following result.
\begin{theorem}
For any two funtions $f , g \in \mathcal{C}_{\mathcal{M}}^{\infty}\left( T^{*}Q \times \mathbb{R}\right)$, we have that,
\begin{enumerate}[i)]
    \item $ X_{H , \mathcal{M}}(f) = X_{H }\left(f  \right)$
    \item $\set{f,g}_{E} = \set{f,g}$
\end{enumerate}
\begin{proof}
Notice that $v \in \orthL{\mathfrak{D}^{l}} = \sharp_{Q}\left(\ann{\mathfrak{D}^{l}}\right) $ if, and only if, there are local functions $\lambda_{a}$ such that,
\begin{equation}\label{ContNH17}
   v\left( q^{i}\right)=0=v\left( z\right), \ \ \  v\left(p_{i}\right) = \lambda_{a} \Phi^{a}_{i}
\end{equation}
Hence, the mechanical condition implies that,
\begin{equation}\label{ContNH16}
 \Phi^{a}_{i} \pdv{f}{p_i} = 0,
\end{equation}
Therefore, we may prove that,
\begin{equation}\label{ContNH18}
    \left(\dd f\right)_{\mathcal{M}\times \mathbb{R}} =  \left(\dd \left(f\circ \gamma\right)\right)_{\mathcal{M}\times \mathbb{R}}
\end{equation}
Then, ${X_{f}}_{\vert \mathcal{M} \times\RR}  = {X_{f \circ \gamma}}_{\vert \mathcal{M} \times\RR}$. Consequently, at elements of $\mathcal{M}\times \mathbb{R}$, we have
$$X_{H , \mathcal{M}}(f) = X_{H }(f \circ \gamma)  = \left\{\left(\dd \left(f\circ \gamma\right)\right) \right\}\left( X_{H}\right)  = \left\{\left(\dd f\right) \right\}\left( X_{H}\right) = X_{H}\left(f \right).$$
The identity $ii)$ follows directly from $i)$.
\end{proof}
\end{theorem}
Therefore, the evolution of observables in $\mathcal{C}_{\mathcal{M}}^{\infty}\left( T^{*}Q \times \mathbb{R}\right)$ within the non-holonomic system coincides with their evolution in the unconstrained system. Furthermore, when restricted to $\mathcal{C}_{\mathcal{M}}^{\infty}\left( T^{*}Q \times \mathbb{R}\right)$, the Eden bracket is identical to the canonical Jacobi bracket of functions in the unconstrained setting. In other words, for any non-holonomic mechanical system, \textit{we have obtained a vector subspace of observables where the dynamics remain unconstrained.}

\section*{Acknowledgments}

M. de Le\'on and V. M. Jiménez acknowledge financial support from the Spanish Ministry of Science and Innovation, under grants PID2022-137909NB-C21 and the Severo Ochoa Program for Centers of Excellence in R$\&$D (CEX2023-001347-S). V. M. Jiménez acknowledge the financial support from MIU and European Union-\textit{NextGenerationUE}. 


\bibliographystyle{plain}

\bibliography{Library}

\end{document}